# High resolution determination of ferromagnetic metallic limit in epitaxial $La_{1-x}Ca_xMnO_3$ films on $NdGaO_3$

D. Sánchez[a], L.E. Hueso[b]

*Department of Materials Science, University of Cambridge, Pembroke Street, Cambridge CB2 3QZ, UK*

L. Granja[c], P. Levy

*Departamento de Física, Comisión Nacional de Energía Atómica, Gral Paz 1499 (1650) San Martín, Buenos Aires, Argentina*

N.D. Mathur[d]

*Department of Materials Science, University of Cambridge, Pembroke Street, Cambridge CB2 3QZ, UK*

The physical properties of manganites depend strongly on sample morphology, probably due to strain. We investigate the influence of $NdGaO_3$ substrates on the limit of the ferromagnetic-metallic phase field in $La_{1-x}Ca_xMnO_3$, doping with $x$=1% resolution. Films with $x$=0.40 show a metal-insulator transition, but the ferromagnetic volume fraction is half the spin aligned value. Films with $x$=0.41 are similar but the metal-insulator transition is not always seen. Films with $x$=0.42, 0.43, 0.44, 0.45 are insulating, and the magnetization is dramatically reduced. The observed phase boundary indicates where to search for enhanced phase separation effects that may be exploited in thin films and devices.

---
[a] Email: ds383@cam.ac.uk
[b] Present address: Spintronic Devices Division, ISMN-CNR, Via P. Gobetti 101, 40129 Bologna, Italia
[c] Also at: Department of Materials Science, University of Cambridge, Pembroke Street, Cambridge CB2 3QZ, UK
[d] Email: ndm12@cam.ac.uk

Manganites such as $La_{1-x}Ca_xMnO_3$ display a rich array of physical phenomena [1,2] such as colossal magnetoresistance (CMR) [3]. These phenomena arise via the interacting magnetic, electronic and crystal structures that give rise to complex phases whose natures may be controversial [4,5]. At present, the most exciting phenomena involve phase coexistence over a wide range of lengthscales [6]. The balance between the competing phases may be tuned using many parameters such as magnetic field and temperature [7-9]. Strain plays a particularly significant role [10-12] and therefore sample morphology will heavily influence physical properties. Epitaxial films have the capacity to reveal basic physical properties because certain extrinsic effects, such as those introduced by grain boundaries, are minimized.

Exploiting phase separation in thin film devices is a challenge that has renewed the interest in the manganites. For example, small driving forces may alter the conductivity leading to memory effects [9,13,14]. In order to optimize the design and performance of phase separated manganite devices, high resolution thin film $x$-$T$ phase diagrams are desirable. In this work, we systematically investigate the limit of ferromagnetism and the percolation threshold in $La_{1-x}Ca_xMnO_3$ thin films with $0.40 \leq x \leq 0.45$, where metallic and insulating phases are likely to coexist.

Series of 20 and 50 nm films of $La_{1-x}Ca_xMnO_3$ were grown on $NdGaO_3$ (001) substrates (NGO) by pulsed laser deposition ($\lambda$=248 nm, 1 Hz, target-substrate distance=6.5 cm, ~800°C, 15 Pa $O_2$, 2 J.cm$^{-2}$) from commercial targets (Praxair, USA) with Ca contents in the range $0.40 \leq x \leq 0.45$ ($\Delta x$=0.01). After deposition, films were annealed in ~50 kPa $O_2$ for 1 hour at the growth temperature, and then cooled quickly to room temperature. X-ray diffraction (XRD) of the targets was performed in a Philips PW1050 diffractometer (CuK$_\alpha$). Films were measured in a High Resolution Philips PW3050/65 X'Pert PRO horizontal diffractometer (CuK$_{\alpha 1}$). The magnetic properties of the target fragments and the films were studied in the cold stage of a Princeton Measurement Corporation vibrating sample magnetometer (VSM). Absolute values of resistivity were determined at room temperature using the van der Pauw technique, with silver dag contacts around the sample perimeter. Variable temperature electrical measurements were carried out in a closed cycle He cryostat using four in-line silver dag contacts.

All targets were found to be single phase within the resolution of our diffractometer, and an orthorhombic *Pnma* unit cell was identified. Lattice parameters were obtained from a Rietveld refinement using the Fullprof program [15]. As expected, the lattice parameters and unit cell volume decrease with increasing Ca content. This indicates that the A-site cation composition displays the nominal monotonic variation within the series of targets. Pseudo-cubic lattice parameters assuming *Pnma* are $a_c^2=(a/2)^2+(c/2)^2$ and $c_c=b/2$. Pseudo-cubic lattice parameters assuming *Pbnm*, typically used for NGO, are $a_c^2=(a/2)^2+(b/2)^2$ and $c_c=c/2$. The in-plane lattice parameters are thus $a_c$=3.864 Å (NGO) [16] and $a_c$=3.855 Å ($La_{1-x}Ca_xMnO_3$, $x$=0.40). The expected lattice mismatch $100 \times (a_c^{LCMO}-a_c^{NGO})/a_c^{LCMO}$ is only -0.2% at $x$=0.40, increasing to -0.5% at $x$=0.45. These values are smaller than for $SrTiO_3$ (-1.3% to -1.5%) or $LaAlO_3$ (1.7% to 1.5%) substrates.

Both 20 nm and 50 nm films were found to be epitaxial and coherently strained. Figure 1 shows representative high resolution XRD scans around the orthorhombic (004) reflection of the NGO substrate indexed as *Pbnm*. The observed thickness fringes confirm both the crystallinity and the coherent nature of the interface. Film thickness as determined from the period of the fringes was 20±2 nm and 50±4 nm for each series. The out of plane lattice parameter of the manganite film could not be calculated with high precision due to overlap of the film peak with the substrate peak as seen in Figure 1.

The magnetic properties of all films are very different from the corresponding targets. All targets are strongly ferromagnetic as expected [17, 18]. However, the saturation magnetization $M_S$ of both 20 and 50 nm films is significantly reduced for $x$=0.40 and 0.41, and too small to be resolved for $x$>0.42. Figure 2 shows the 50 K hysteresis loops for 20 nm films with $x$=0.40, 0.41 and 0.42, after substracting the paramagnetic NGO substrate contribution. $M_S$ values are 1.71 $\mu_B$/Mn, 1.43 $\mu_B$/Mn and 0.14 $\mu_B$/Mn, respectively. The dramatic reduction seen at $x$=0.42 is identified as a thin film effect because it is not seen in the corresponding target where $M_S$=3.10 $\mu_B$/Mn (Figure 2, inset).

Table I gives ferromagnetic volume fractions for the 20 and 50 nm films at $x$=0.40, 0.41 and 0.42. Due to the strongly paramagnetic substrate, the existence of ferromagnetism in these films at high temperatures cannot be reliably resolved, and therefore Curie temperatures are not presented. Similarly, we cannot resolve a ferromagnetic component in films with $x$=0.43, 0.44 and 0.45.

Figure 3 shows selected resistivity-temperature data for the 20 nm film series, which is qualitatively identical for the 50 nm film series. A metal-insulator transition is always seen in $x$=0.40 films, but only sometimes in films with $x$=0.41. Films studied with $x \geq 0.42$ are always insulating in the measured temperature range. By contrast, all targets studied shows the expected [17, 18] metal-insulator transition (see e.g. Figure 3, inset).

Given that high quality films with $x$=0.30 have been grown in this laboratory under equivalent conditions [20], the magnetic and electrical data presented above may be attributed to thin film effects. We summarize and interpret our results as follows. Films at $x$=0.40 are qualitatively similar to films that lie deep within the ferromagnetic metallic phase around $x$=3/8 [17, 18], but the ferromagnetic phase fraction is halved (Figure 2). Therefore the resistivity is increased and the metal-insulator transition is suppressed. Films with $x$=0.41 always display some degree of ferromagnetism but not every sample displays metallicity. This suggests that $x$=0.41 films are also phase separated, but unlike $x$=0.40 films they lie near a percolation threshold such that their physical properties are very susceptible to subtle variations, e.g. in strain. The ferromagnetic metallic state is dramatically suppressed in films with $x \geq 0.42$ but a very small phase fraction may persist.

Based on the above findings, a thin film phase diagram for epitaxial $La_{1-x}Ca_xMnO_3$ on $NdGaO_3$ is presented in Figure 4. It differs dramatically from bulk phase diagram [17,18] because phase separation halves the phase fraction of the ferromagnetic metal, and reduces the limit of metallicity and ferromagnetism by $x \approx 0.09$ from $x$=0.5 in the bulk to $x \approx 0.41$ in our films. Given that oxygen deficiencies would tend to electron dope $La_{1-x}Ca_xMnO_3$ and therefore increase rather than reduce the limit of metallicity and ferromagnetism, we attribute our observations to epitaxial strain. The apparent percolation threshold near $x=0.41$ could be exploited to optimize e.g. memory effects [9,13,14] or low-field magnetoresisance [23,24] in devices based upon phase separated manganites.

Financial support by the UK EPSRC, the EU, The Royal Society and ANPCYT PICT No. 03-13517 (Argentina) is acknowledged. The authors thank E.C. Israel and H.Y. Hwang for helpful comments.

**Table I.** Ferromagnetic volume fractions in the 20 nm and 50 nm film series for $x$=0.40, 0.41 and 0.42. Values calculated from dividing the saturation magnetization $M_S$ by the fully spin-aligned value (4-$x$) $\mu_B$/Mn.

| Thickness | $x$ = 0.40 | $x$ = 0.41 | $x$ = 0.42 |
|---|---|---|---|
| 20 nm | 0.47 | 0.38 | 0.05 |
| 50 nm | 0.54 | 0.47 | 0.17 |

**Figure Captions**

Figure 1. High-resolution XRD ω-2θ scans of representative films in the series $La_{1-x}Ca_xMnO_3$ $0.40 \leq x \leq 0.45$. The sharp substrate (004) peak overlaps with the corresponding film peak, and thickness fringes may be seen.

Figure 2. 50 K magnetization *M* versus field *B* loops for 20 nm thick films with *x*=0.40, 0.41 and 0.42. The magnetometer axis was parallel to the orthorhombic [100] film easy axis [19]. Inset: corresponding loop for the *x*=0.42 target at 50 K.

Figure 3. Resistivity $\rho$ versus temperature *T* for 20 nm films with *x*=0.40, 0.41 and 0.42. Inset: resistance *R* versus *T* for the *x*=0.42 target. The small suppression in the transition temperature may be due to deoxygenation.

Figure 4. Thin film phase diagram for 20 nm epitaxial $La_{1-x}Ca_xMnO_3$ on NGO in $0.40 \leq x \leq 0.45$. Metal-insulator transition temperatures (Δ) and 50 K ferromagnetic phase fractions (•) are plotted. The straight lines are a guide to the eye. The boundary between the phase separated metallic phase (PSM) and the paramagnetic insulating (PMI) phase assumes that percolation requires a ferromagnetic fraction of 10%, which is a lower bound [21]. The inset shows the corresponding phase diagram for polycrystalline samples (data from [18, 22]) where the phase fraction of the ferromagnetic metallic phase (FMM) is ~0.9 for x = 0.45 [18]. Open triangles (Δ) correspond to metal-insulator transition temperatures.

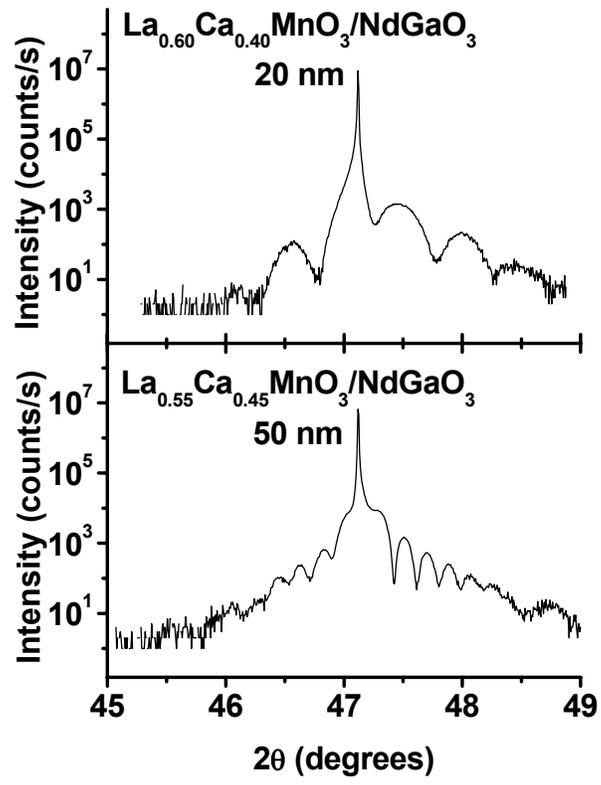

**Figure 1**

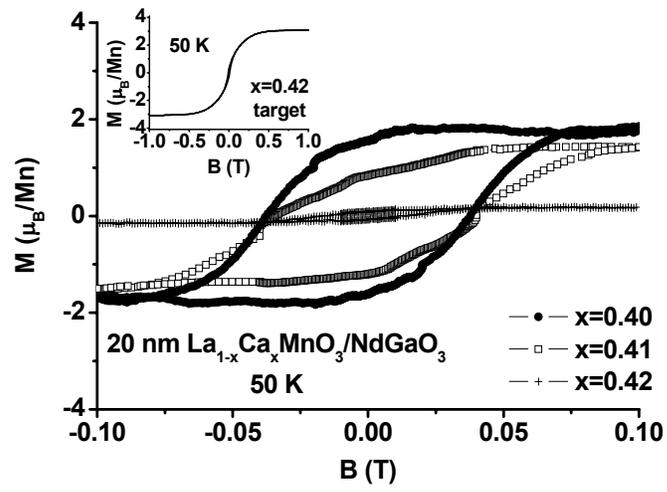

**Figure 2**

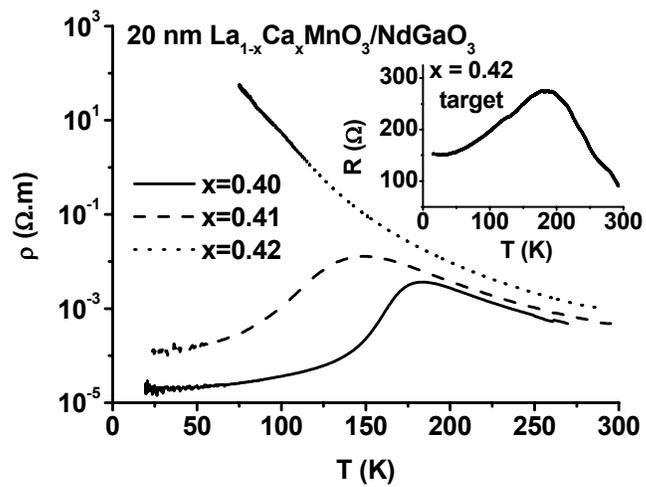

**Figure 3**

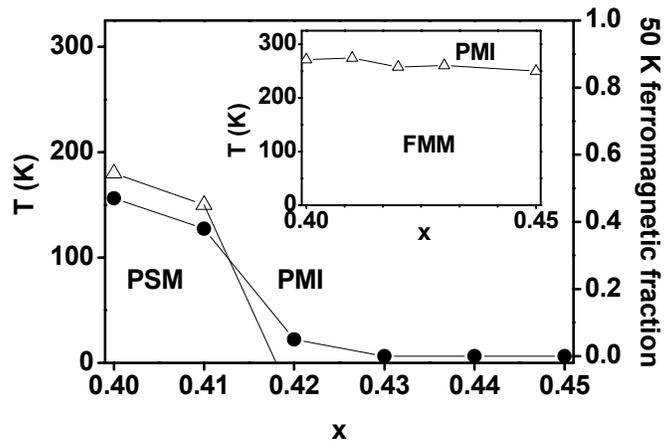

**Figure 4**